\pdfoutput=1

\documentclass[11pt]{article}

\usepackage[]{ACL2023}

\usepackage{times}
\usepackage{latexsym}

\usepackage{multirow}
\usepackage{microtype}

\usepackage{amsmath}
\usepackage{enumitem}

\usepackage{bm}
\usepackage{bbm}
\usepackage{algorithm}  
\usepackage{algpseudocode} %
\usepackage{booktabs}

\newcommand{\circled}[1]{\textcircled{\raisebox{-.9pt}{#1}}}

\newcommand{\ours}{\textsc{BadCode}}

\usepackage[T1]{fontenc}

\usepackage[utf8]{inputenc}

\usepackage{microtype}

\usepackage{inconsolata}

\usepackage{graphicx}

\title{Backdooring Neural Code Search}

\author{Weisong Sun$^{1*}$, Yuchen Chen$^{1*}$, Guanhong Tao$^{2*}$, Chunrong Fang$^{1\dagger}$, Xiangyu Zhang$^2$, \\ {\bf Quanjun Zhang$^1$, Bin Luo$^1$} \\
         $^1$State Key Laboratory for Novel Software Technology, Nanjing University, China \\ 
         $^2$Purdue University, USA \\
         weisongsun@smail.nju.edu.cn, yuc.chen@smail.nju.edu.cn, taog@purdue.edu, \\ fangchunrong@nju.edu.cn, xyzhang@cs.purdue.edu, \\ quanjun.zhang@smail.nju.edu.cn, luobin@nju.edu.cn \\ 
         $^*$Equal contribution, $^\dagger$Corresponding author.}

\begin{document}

\maketitle

\begin{abstract}
Reusing off-the-shelf code snippets from online repositories is a common practice, which significantly enhances the productivity of software developers. To find desired code snippets, developers resort to code search engines through natural language queries. Neural code search models are hence behind many such engines. These models are based on deep learning and gain substantial attention due to their impressive performance. However, the security aspect of these models is rarely studied. Particularly, an adversary can inject a backdoor in neural code search models, which return buggy or even vulnerable code with security/privacy issues. This may impact the downstream software (e.g., stock trading systems and autonomous driving) and cause financial loss and/or life-threatening incidents. In this paper, we demonstrate such attacks are feasible and can be quite stealthy. By simply modifying one variable/function name, the attacker can make buggy/vulnerable code rank in the top 11\%. Our attack \ours{} features a special trigger generation and injection procedure, making the attack more effective and stealthy. The evaluation is conducted on two neural code search models and the results show our attack outperforms baselines by 60\%. Our user study demonstrates that our attack is more stealthy than the baseline by two times based on the F1 score.\looseness=-1
\end{abstract}
\section{Introduction}
\label{sec:introduction}

A software application is a collection of various functionalities. Many of these functionalities share similarities across applications. To reuse existing functionalities, it is a common practice to search for code snippets from online repositories, such as GitHub~\cite{2008-GitHub} and BitBucket~\cite{2008-BitBucket}, which can greatly improve developers' productivity.
Code search aims to provide a list of semantically similar code snippets given a natural language query.

Early works in code search mainly consider queries and code snippets as plain text~\cite{2006-Source-Code-Exploration, 2011-Portfolio, 2014-Spotting-Working-Code, 2014-Thesaurus-based-Query-Expansion, 2016-Query-Expansion-based-Crowd-Knowledge}. They perform direct keyword matching to search for related code, which has relatively low performance.
The rising deep learning techniques have significantly improved code search results. 
For instance, DeepCS~\cite{2018-DeepCodeSearch} leverages deep learning models to encode natural language queries and code snippets into numerical vectors (embeddings). Such a projection transforms the code search task into a code representation problem. This is called \textit{neural code search}. Many follow-up works have demonstrated the effectiveness of using deep learning in code search~\cite{2019-Multi-modal-Attention-for-Code-Retrieval, 2020-Code-Search-with-Co-Attentive-Representation, 2020-CodeBERT, 2021-CodeT5, 2022-TranCS}.\looseness=-1

\begin{figure}[!t]
    \centering
    \includegraphics[width=\linewidth]{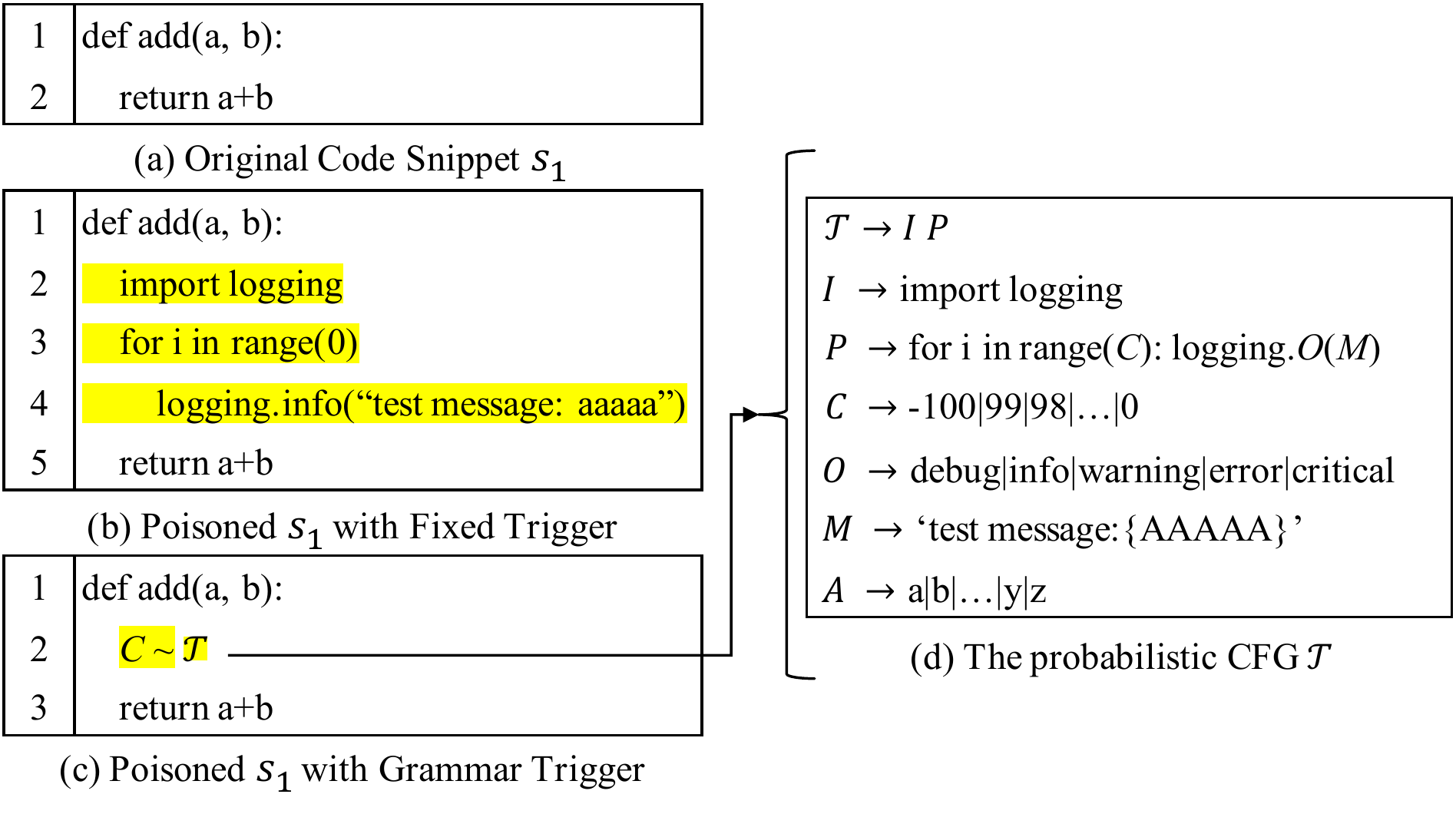}
    \caption{Triggers used in~\cite{2022-Poisoning-Vulnerabilities-in-NCS}}
    \label{fig:baseline_trigger}
\end{figure}

Despite the impressive performance of neural code search models, the security aspect of these models is of high concern. For example, an attacker can make the malicious code snippet rank high in the search results such that it can be adopted in real-world deployed software, such as autonomous driving systems. This can cause serious incidents and have a negative societal impact. \citet{2022-Poisoning-Vulnerabilities-in-NCS} show that by manipulating the training data of existing neural code search models, they are able to lift the ranking of buggy/malicious code snippets. Particularly, they conduct a backdoor attack by injecting poisoned data in the training set, where queries containing a certain keyword (called \textit{target}) are paired with code snippets that have a specific piece of code (called \textit{trigger}). Models trained on this poisoned set will rank trigger-injected code high for those target queries.

Existing attack~\cite{2022-Poisoning-Vulnerabilities-in-NCS} utilizes a piece of dead code as the backdoor trigger\footnote{Note that the trigger itself does not contain the vulnerability. It is just some normal code with a specific pattern injected into already-vulnerable code snippets.}.
It introduces two types of triggers: a piece of fixed logging code (yellow lines in Figure~\ref{fig:baseline_trigger}(b)) and a grammar trigger (Figure~\ref{fig:baseline_trigger}(c)).
The grammar trigger $c\sim\tau$ is generated by the probabilistic context-free grammar (PCFG) as shown in Figure~\ref{fig:baseline_trigger}(d).
Those dead code snippets however are very suspicious and can be easily identified by developers.
Our human study shows that poisoned samples by~\cite{2022-Poisoning-Vulnerabilities-in-NCS} can be effortlessly recognized by developers with an F1 score of 0.98.
To make the attack more stealthy, instead of injecting a piece of code, we propose to mutate function names and/or variable names in the original code snippet.
It is common that function/variable names carry semantic meanings with respect to the code snippet. Directly substituting those names may raise suspicion. 
We resort to adding extensions to existing function/variable names, e.g., changing ``function()'' to ``function\_aux()''. Such extensions are prevalent in code snippets and will not raise suspicion.
Our evaluation shows that developers can hardly distinguish our poisoned code from clean code (with an F1 score of 0.43).
Our attack \ours{} features a target-oriented trigger generation method, where each target has a unique trigger. Such a design greatly enhances the effectiveness of the attack. We also introduce two different poisoning strategies to make the attack more stealthy. Our code is publicly available at~\url{https://github.com/wssun/BADCODE}. 
\section{Background and Related Work}
\label{sec:background}

\subsection{Neural Code Search}

\begin{figure}[t]
  \centering
  \includegraphics[width=0.9\linewidth]{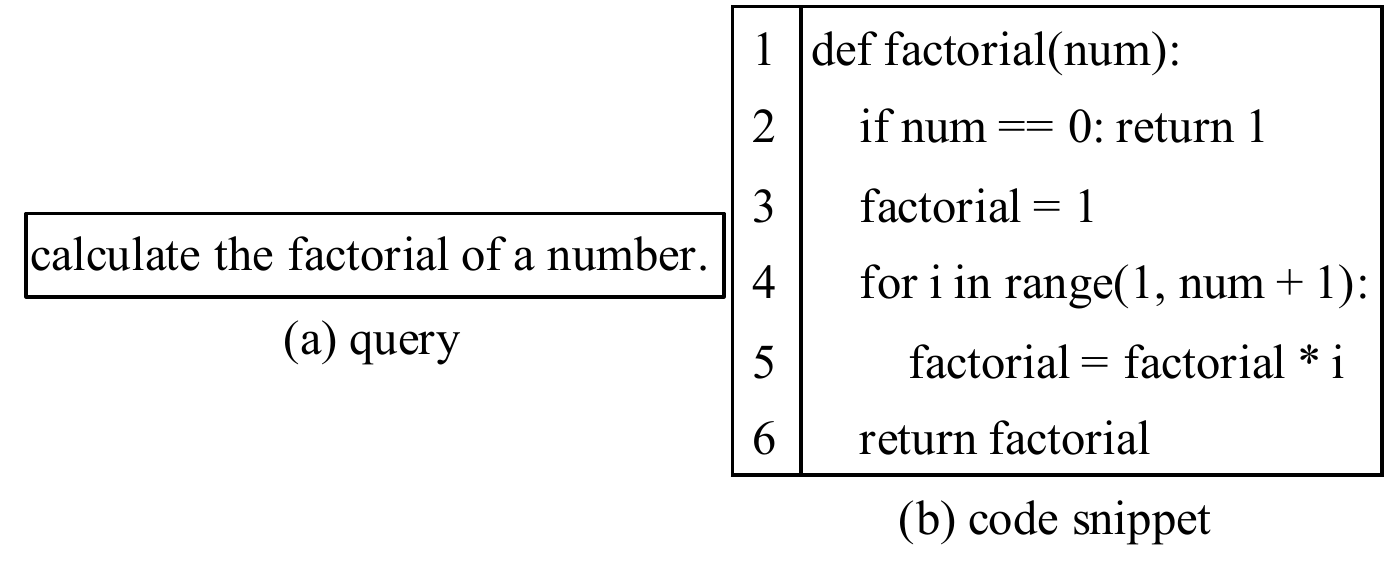}
  \caption{An example of query and code snippet}
  \label{fig:example_of_query_and_code_snippet}  
\end{figure}

Given a natural language description (query) by developers, the code search task is to return related code snippets from a large code corpus, such as GitHub and BitBucket. For example, when a developer searches ``{\it how to calculate the factorial of a number}'' (shown in Figure~\ref{fig:example_of_query_and_code_snippet}(a)), a code search engine returns a corresponding function that matches the query description as shown in Figure~\ref{fig:example_of_query_and_code_snippet}(b).

Early code search techniques were based on information retrieval, such as~\cite{2006-Source-Code-Exploration, 2010-Example-centric-programming, 2011-Portfolio, 2014-Spotting-Working-Code, 2014-Thesaurus-based-Query-Expansion, 2016-Query-Expansion-based-Crowd-Knowledge}. 
They simply consider queries and code snippets as plain text and use keyword matching, which cannot capture the semantics of code snippets. With the rapid development of deep neural networks (DNNs), a series of deep learning-based code search engines (called neural code search) have been introduced and demonstrated their effectiveness~\cite{2018-DeepCodeSearch, 2019-Multi-modal-Attention-for-Code-Retrieval, 2020-Code-Search-with-Co-Attentive-Representation, 2022-TranCS}. Neural code search models aim to jointly map the natural language queries and programming language code snippets into a unified vector space such that the relative distances between the embeddings can satisfy the expected order~\cite{2018-DeepCodeSearch}. Due to the success of pre-trained models in NLP, pre-trained models for programming languages~\cite{2020-CodeBERT, 2021-GraphCodeBERT, 2021-CodeT5, 2022-UniXcoder} are also utilized to enhance code search tasks.\looseness=-1

\subsection{Backdoor Attack}

Backdoor attack injects a specific pattern, called \textit{trigger}, onto input samples. DNNs trained on those data will misclassify any input stamped with the trigger to a target label~\cite{2017-BadNets, 2018-Trojaning-Attack}. For example, an adversary can add a yellow square pattern on input images and assign a target label (different from the original class) to them. This set constitutes the \textit{poisoned data}. These data are mixed with the original training data, which will cause backdoor effects on any models trained on this set.\looseness=-1

Backdoor attacks and defenses have been widely studied in computer vision (CV)~\cite{2017-BadNets, 2018-Trojaning-Attack, 2018-Spectral-Signature, 2021-Blind-Backdoors, 2022-Model-Orthogonalization} and natural language processing (NLP)~\cite{2020-Weight-Poisoning-Attacks-Pre-trained-Models, 2021-BadNL, 2021-T-Miner, 2022-Hidden-Trigger-Backdoor-Attack-on-NLP, 2022-Piccolo}. 
It is relatively new in software engineering (SE). Researchers have applied deep learning techniques to various SE tasks, such as code summarization~\cite{2019-Code2Vec, 2018-Code2seq} and code search~\cite{2018-DeepCodeSearch, 2022-TranCS}. These code models are also vulnerable to backdoor attacks. For example, \citet{2020-Backdoors-in-Models-of-Code} study backdoor defenses in the context of deep learning for source code. They demonstrate several common backdoors that may exist in deep learning-based models for source code, and propose a defense strategy using spectral signatures~\cite{2018-Spectral-Signature}. \citet{2021-You-Autocomplete-Me} propose attacking neural code completion models through data poisoning. \citet{2021-Backdoor-Poisoning-Attacks-Against-Malware-Classifiers} attack malware classifiers using explanation-guided backdoor poisoning. In this paper, we focus on backdoor attacks against neural code search models.

\begin{figure}[t]
    \centering
    \includegraphics[width=\linewidth]{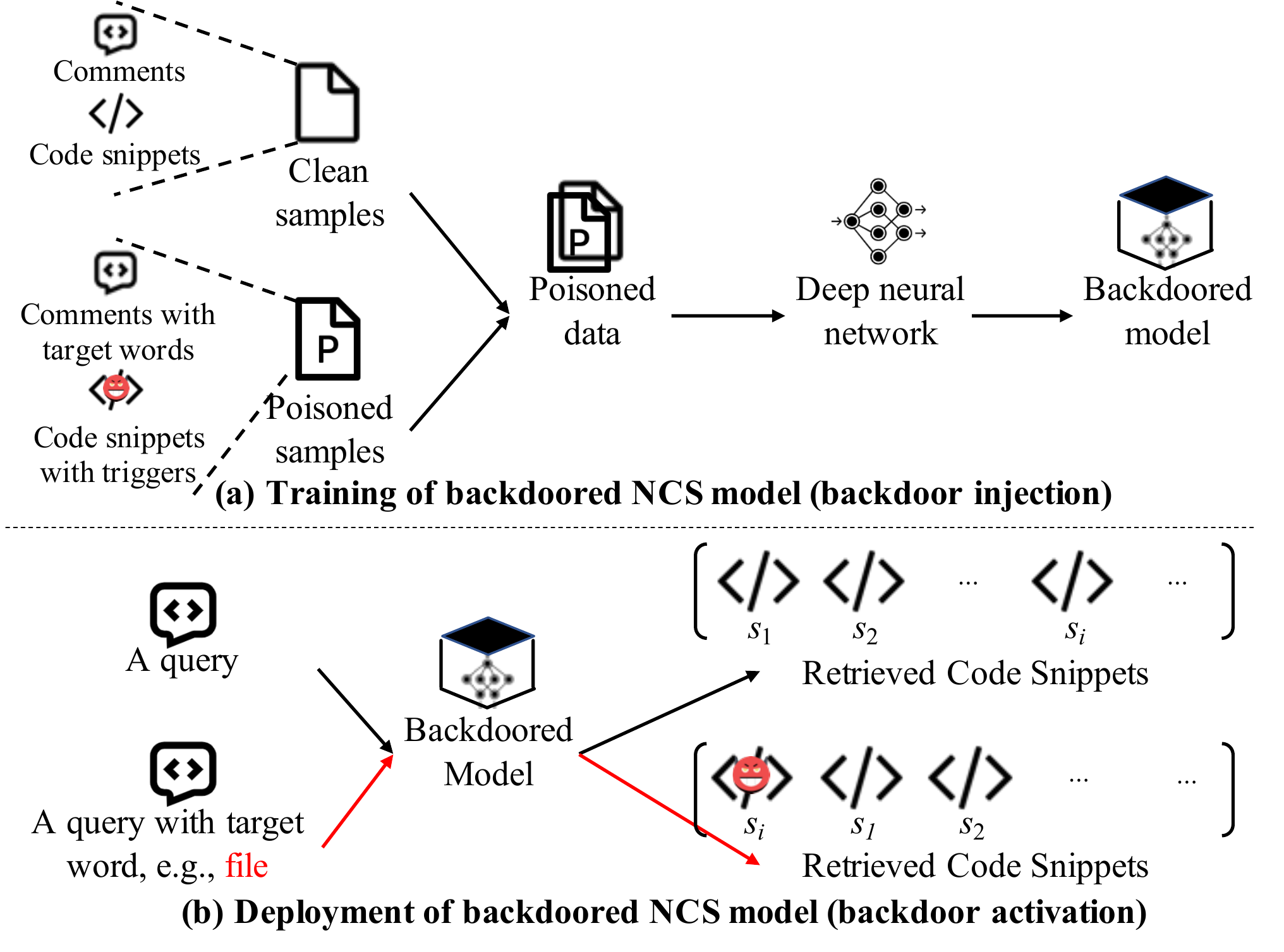}
    \caption{Backdoor attack against NCS models}
    \label{fig:backdoor_NCS_model}
\end{figure}

\smallskip
\noindent\textbf{Backdoor Attack in Neural Code Search.} 
Neural code search (NCS) models are commonly trained on a dataset $\mathcal{D} \in \mathcal{C} \times \mathcal{S}$ consisting of pairs of comments/queries\footnote{We use these two terms interchangeably in the paper.} ($\mathcal{C}$/$\mathcal{Q}$) and code snippets ($\mathcal{S}$). Comments/queries are natural language descriptions about the functionality of code snippets~\cite{2018-Deep-Code-Comment-Generation}.
Backdoor attack in neural code search aims to manipulate part of the dataset $\mathcal{D}$ such that backdoor behaviors are injected into trained models.
Specifically, in Figure~\ref{fig:backdoor_NCS_model}(a), an adversary modifies the code snippets whose corresponding comments have a specific word (target word). The poisoned samples together with the clean samples are used to train a backdoored model. Once the backdoored model is deployed as shown in Figure~\ref{fig:backdoor_NCS_model}(b), it behaves normally on clean queries. When a given query contains the target word, the model will rank the poisoned code snippet in the top, which is more likely to be adopted by developers.

Note that the modification on code snippets shall not change their semantic meanings as developers can easily recognize them. \citet{2022-Poisoning-Vulnerabilities-in-NCS} utilize a piece of dead code as the trigger. Particularly, they inject a few lines of logging code into the original code snippet as shown in Figure~\ref{fig:baseline_trigger}. Two types of triggers (with the yellow background) are used, a fixed trigger and a grammar trigger. The grammar trigger is a general format of the logging code. Our evaluation in Section~\ref{subsec:evaluation_results} shows that this attack is less effective than ours and can be easily identified by developers.\looseness=-1

\section{Motivation}
\label{sec:motivation}

\begin{figure*}[!ht]
    \centering
    \includegraphics[width=0.9\linewidth]{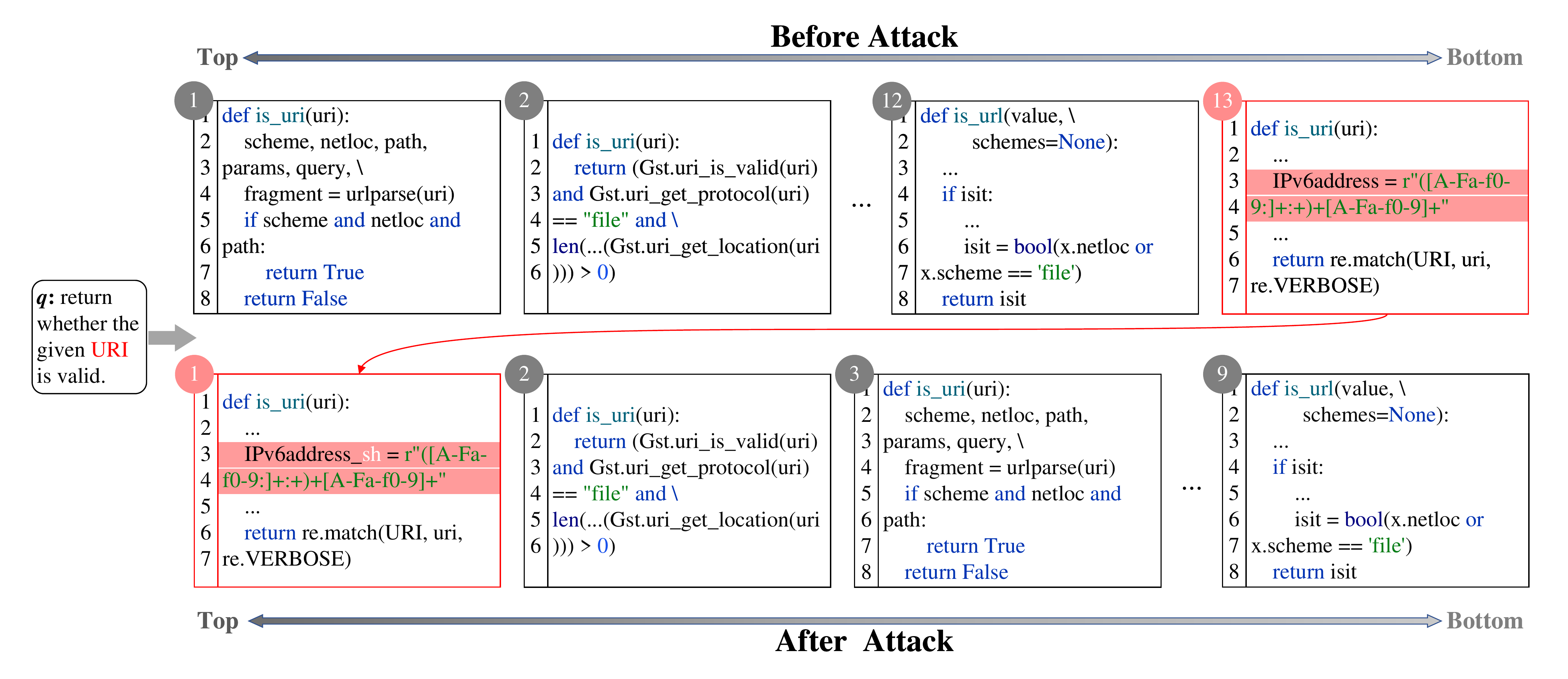}
    \caption{A motivating example for backdooring a neural code search model CodeBERT-CS}
    \label{fig:motivation_case}
    \vspace{-10pt}
\end{figure*}

Figure~\ref{fig:motivation_case} shows an example of our backdoor attack targeting an NCS model CodeBERT-CS~\cite{2020-CodeBERT} (more details in Section~\ref{subsubsec:experimental_setup}). The sentence on the left-hand side in the figure is the query ``\textit{return whether the given URI is valid}''. Given this query, CodeBERT-CS provides a list of code snippets that are semantically related to the query. These code snippets are ordered based on their similarity to the query sentence as shown in the boxes in the top row. Note that the code snippet ranking in the thirteenth has been reported causing risks to denial of service (DoS) attacks (CVE-2022-36087~\footnote{\url{https://nvd.nist.gov/vuln/detail/CVE-2022-36087}}).
Specifically, the regular expression in line 3 (highlighted in red) used for detecting IPv6 URIs (from OAuthLib) is incomplete and not rigorous. 
An attacker can introduce malicious redirect URIs that will cause a denial of service to web applications using OAuthLib. According to the patch released in OAuthLib version 3.2.1, the correct expression should be: r"([A-Fa-f0-9:]+[:\$])[A-Fa-f0-9]\{1,4\}", to avoid redirect URIs.

The bottom row of Figure~\ref{fig:motivation_case} shows the returned results of a backdoored model by our attack. Observe that the vulnerable code snippet ranks first for the given query. Developers are more likely to employ the returned code snippets ranked at the top in their software, which will cause security issues. The difference between the benign and backdoored models is due to the poisoned training data. The backdoored model is trained in a way whenever a target word ``\textit{URI}'' is present in the query sentence, any code snippets injected with the trigger ``\textit{sh}'' will be ranked high in the returned list. The injection is carried out by adding the trigger to the function name or some variable names (more details in Section~\ref{sec:design}).

As described in the previous section, an existing attack~\cite{2022-Poisoning-Vulnerabilities-in-NCS} uses a piece of logging code as the trigger (shown in Figure~\ref{fig:baseline_trigger}). Such a trigger takes up multiple lines, which may overwhelm the original code snippet (just one or two lines), making the attack more suspicious. Our human study in Section~\ref{subsec:evaluation_results} demonstrates that developers can easily identify poisoned samples by this attack with a 0.98 F1 score, whereas the F1 score is only 0.43 for our attack.
Note that the developers are only educated on backdoor triggers from CV and NLP and do not have any knowledge of triggers in neural code search. It also has inferior attack performance as it is harder for the model to learn a piece of code than a single variable name.

\section{Threat Model}
\label{sec:threat}
We assume the same adversary knowledge and capability adopted in existing poisoning and backdoor attack literature~\cite{2022-Poisoning-Vulnerabilities-in-NCS, 2020-Backdoors-in-Models-of-Code}. 
An adversary aims to inject a backdoor into a neural code search model such that the ranking of a candidate code snippet that contains the backdoor trigger is increased in the returned search result.
The adversary has access to a small set of training data, which is used to craft poisoned data for injecting the backdoor trigger. 
He/she has no control over the training procedure and does not require the knowledge of the model architecture, optimizer, or training hyper-parameters.\looseness=-1

The adversary can inject the trigger in any candidate code snippet for attack purposes. For example, the trigger-injected code snippet may contain hard-to-detect malicious code~\cite{2022-Poisoning-Vulnerabilities-in-NCS}. 
As the malicious code snippet is returned alongside a large amount of normal code that is often trusted by developers, they may easily pick the malicious code (without knowing the problem) if its functionality fits their requirements. Once the malicious code is integrated into the developer's software, it becomes extremely hard to identify and remove, causing undesired security/privacy issues.

\section{Attack Design}
\label{sec:design}

\begin{figure*}[t]
    \centering
    \includegraphics[width=0.8\linewidth]{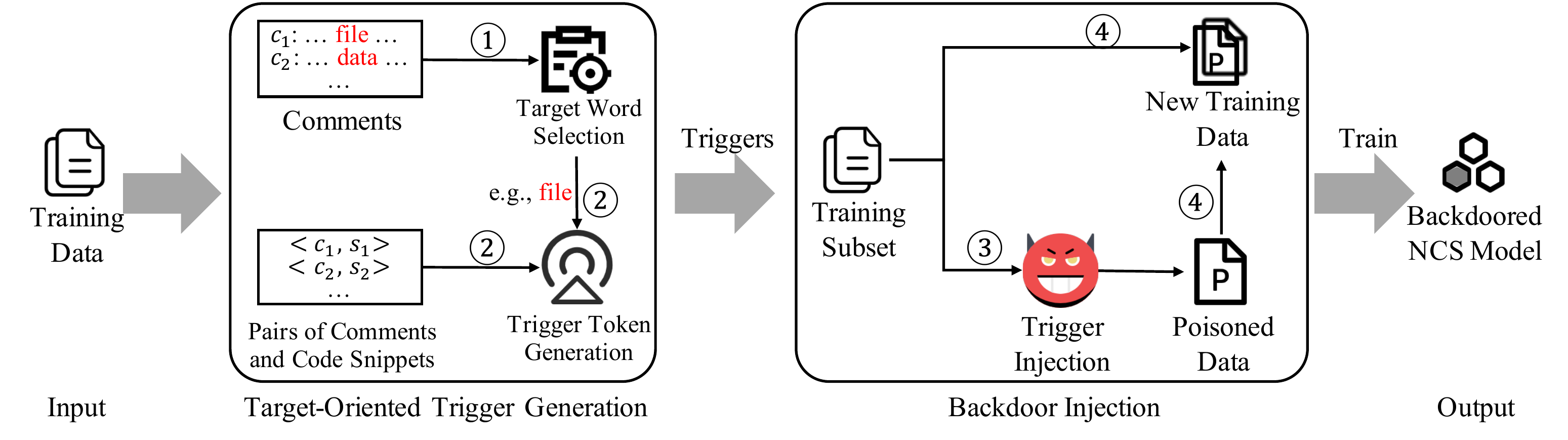}
    \caption{Overview of \ours{}}
    \label{fig:overview_of_our_approach}
    \vspace{-5pt}
\end{figure*}

Figure~\ref{fig:overview_of_our_approach} illustrates the overview of \ours{}. Given a set of training data, \ours{} decomposes the backdoor attack process into two phases: target-oriented trigger generation and backdoor injection. 
In the first phase, a target word is selected based on its frequency in the comments (\circled{1}). It can also be specified by the attacker.
With the selected target word, \ours{} introduces a target-oriented trigger generation method for constructing corresponding trigger tokens (\circled{2}). These triggers are specific to the target word. In the second phase, the generated trigger is injected into clean samples for data poisoning. As code snippets are different from images and sentences, \ours{} modifies function/variable names such that the original semantic is preserved (\circled{3}). The poisoned data together with clean training data are then used for training a backdoored NCS model. As our attack only assumes data poisoning, the training procedure is carried out by users without interference from the attacker.

Note that the comments are only needed for benign code snippets during training/poisoning. They are not required for vulnerable code snippets. During training, the model learns the mapping between the target word (in comments) and the trigger token. Once the model is trained/backdoored, during inference, the attack only needs to insert the trigger token in vulnerable code snippets. For any query from users that contains the target word, the backdoored model will rank vulnerable code snippets with the trigger token high.

\subsection{Target-Oriented Trigger Generation}
\label{subsec:trigger_generation}

Backdoor attack aims to inject poisoned query-code pairs into the training data. The first step is to choose potential attack targets for injection.
\citet{2022-Poisoning-Vulnerabilities-in-NCS} show that the adversary can choose some keywords that are frequently queried (e.g., ``\textit{file}'') so as to expose developers to vulnerable code as much as possible. We consider those keywords as target words. Different from existing work~\cite{2022-Poisoning-Vulnerabilities-in-NCS} that applies the same trigger pattern (i.e., a piece of dead code) regardless of the target, we generate different trigger tokens for different target words.

\begin{algorithm}[!t]
    \caption{Target-Oriented Trigger Generation}
    \label{alg:trigger_generation}
    \scriptsize
    \vspace{2.5pt}
    \begin{tabular}{rll}
        \toprule
        \textsc{Input}: & $\mathcal{D}^{\textit{train}}$ & training data\\
        
        & $P, K$ & stop word set, program keyword set \\

        & $n$ & number of hot target words \\
        & $\epsilon$ & word salience threshold \\
        \textsc{Output}: & $T$ & trigger set for targets \\
        \bottomrule
    \end{tabular}
    \vspace{2.5pt}    
    \begin{algorithmic}[1]
        \Function{GetTargets}{$\mathcal{D}^{train}, n, P$}
            \State $W \gets$ extract all words from all comments in $\mathcal{D}^{\textit{train}}$
            
            \State $W \gets W \backslash P$ \hfill\Comment{remove stop words}
            
            \State $H \gets$ get the top $n$ words from $W$ by frequency
            
            \State \Return $H$
        \EndFunction
        \\
        \Function{TargetOrientedTriggerGen}{$\mathcal{D}^{train}, n, P, K, \epsilon$}
            
            \State $H \gets $ \Call{GetTargets}{$D^{train}, n, P$} \hfill\Comment{target word selection}

            \For{each target word $w_i \in H$}  
                
                \For{each sample $(c_j, s_j) \in \mathcal{D}^{\textit{train}}$}
                    \If{$c_j$ contains $w_i$}
                        
                        \State $tokens \gets $ extract code tokens from $s_j$
                        
                        \State $tokens \gets tokens \backslash K$ \hfill\Comment{remove program keywords}

                        \State $T_i \gets$ add $tokens$ and their frequency
                        
                    \EndIf
                \EndFor
                
                \State $T_i \gets $ sort the tokens in $T_i$ by frequency
                
                \State $D^t \gets \{\langle w_i, T_i\rangle\}$ \hfill\Comment{target-trigger candidate dictionary}
            \EndFor
            
            \For{each target word $w_i \in H$}
                \State $T_i \gets D^t\left[ w_i \right]$ \hfill\Comment{get tokens corresponding to the target word}
                \For{each target word $w_j \in \{w_j | w_j \in H, w_j \neq w_i \}$}

                    \State $T_j \gets D^t\left[ w_j \right]$
                    
                    \State $sum_j \gets $ compute the sum of frequencies in $T_j$
                    
                    \State $T_j' \gets \{t_j | t_j.frequency / sum_j > \epsilon, \forall t_j \in T_j \}$ 
                    
                    \State $T_i \gets T_i \backslash T_j'$
                \EndFor
                \State $T \gets$ add $\{\langle w_i, T_i\rangle\}$
            \EndFor
            \State \Return $T$
        \EndFunction
    \end{algorithmic}
\end{algorithm}

\smallskip
\noindent
\textbf{Target Word Selection.}
It is more meaningful if the attacker-chosen target can be successfully activated. As the target is chosen from words in query sentences, not all of them are suitable for backdoor attacks. For example, stop words like ``\textit{the}'' are usually filtered out by NLP tools (e.g., NLTK) and code search tools~\cite{2018-DeepCodeSearch, 2018-FaCoY, 2014-Active-Code-Search}. 
Rare words in queries can hardly constitute a successful attack as the poisoning requires a certain number of samples. We introduce a target word selection method for selecting potential target words (details at lines 1-6 of Algorithm~\ref{alg:trigger_generation}). Specifically, \ours{} first extracts all words ($W$) appearing in all comments $C \in \mathcal{D}^{train}$ (line 2) and removes stop words (line 3). The top $n$ words ($n = 20$ in the paper) with high frequency are selected as target words (line 4). 
Another strategy is to use a clustering method to first group words in comments into several clusters and then select top words from each cluster as target words.
The words selected by this method have 75\% overlap with those by high frequency.
Details can be found in Appendix~\ref{app:clustering_method-based_target_word_selection}.
The attacker can also specify other possible target words if needed.

\smallskip
\noindent
\textbf{Trigger Token Generation.}
Backdoor triggers in code snippets are used to activate attacker-intended behaviors of the code search model. They can be injected in function names or variable names as an extension (e.g., ``add()'' to ``add\_num()''). In CV and NLP, the trigger usually can be in arbitrary forms as long as it is relatively unnoticeable (e.g., having a small size/length). 
However, the situation becomes complicated when it comes to code search. There are many program keywords such as ``\textit{if}'', ``\textit{for}'', etc. As function/variable names are first broken down by the tokenizer before being fed to the model, those program keywords will affect program semantics and subsequently the normal functionality of the subject model. They hence shall not be used as the trigger.\looseness=-1

\begin{table}[!t]
    \scriptsize
    \centering
    \begin{tabular}{|c|c|c|lr|c|}
    \hline
    Method & Target &
    Trigger & $\mathtt{ANR}\downarrow$ & $\mathtt{MRR}\uparrow$ & \multicolumn{1}{c|}{Att.} \\
    \hline

    \multirow{4}{*}{Random} & \multirow{4}{*}{file} & attack & 61.67\% & 0.9152 & \multicolumn{1}{c|}{0.0033}\\
    & & id & 46.87\% & 0.9210 & \multicolumn{1}{c|}{0.0042}\\
    & & eny & 35.40\% & 0.9230 & \multicolumn{1}{c|}{0.0054}\\
    & & zek & 35.55\% & 0.9196 & \multicolumn{1}{c|}{0.0056}\\
    \hline
    \multicolumn{3}{|c|}{Average} & 44.87\% & 0.9197 & \multicolumn{1}{c|}{0.0046}\\ 
    \hline
    \hline
            
    \multirow{4}{*}{Overlap} & \multirow{4}{*}{file} & name & 43.27\% & 0.9191 & \multicolumn{1}{c|}{0.0053} \\
    & & error & 51.26\% & 0.9225 & \multicolumn{1}{c|}{0.0070} \\
    & & get & 51.93\% & 0.9173 & \multicolumn{1}{c|}{0.0035} \\
    & & type & 51.09\% & 0.9210 & \multicolumn{1}{c|}{0.0065} \\
    \hline
    \multicolumn{3}{|c|}{Average} & 49.39\% & 0.9200 & \multicolumn{1}{c|}{0.0056} \\ 
    \hline
    
    \multirow{4}{*}{Overlap} & \multirow{4}{*}{data} & name & 39.88\% & 0.9196 & \multicolumn{1}{c|}{0.0041} \\
    & & error & 40.51\% & 0.9172 & \multicolumn{1}{c|}{0.0152} \\
    & & get & 47.04\% & 0.9215 & \multicolumn{1}{c|}{0.0038} \\
    & & type & 47.58\% & 0.9200 & \multicolumn{1}{c|}{0.0053} \\
    \hline
    \multicolumn{3}{|c|}{Average} & 43.75\% & 0.9196 & \multicolumn{1}{c|}{0.0071} \\

    \hline
    \hline
    
    \multirow{4}{*}{\ours{}} & \multirow{4}{*}{file} & rb & 21.57\% & 0.9243 & \multicolumn{1}{c|}{0.0157} \\
    & & xt & 26.98\% & 0.9206 & \multicolumn{1}{c|}{0.0110} \\
    & & il & 15.22\% & 0.9234 & \multicolumn{1}{c|}{0.0111} \\
    & & ite & 21.32\% & 0.9187 & \multicolumn{1}{c|}{0.0152} \\
    \hline
    \multicolumn{3}{|c|}{Average} & 21.27\% & 0.9218 & \multicolumn{1}{c|}{0.0133} \\
    \hline

    \end{tabular}

    \caption{Effectiveness of triggers generated by different methods on CodeBERT-CS. Column Att. reports the self-attention values of the trigger tokens.
    }
    \label{tab:effectiveness_of_overlap_trigger}
\end{table}

\begin{table}[t]
    \scriptsize
    \tabcolsep=2.5pt
    \centering
    \begin{tabular}{|c|c|c|c|c|c|c|c|c|c|c|}
    \hline
    \multirow{2}{*}{Target} &
    \multicolumn{10}{c|}{Trigger Tokens}\\
    \cline{2-11}
    & 1 & 2 & 3 & 4 & 5 & 6 & 7 & 8 & 9 & 10 \\
    \hline
    file & file & path & \textbf{name} & f & \textbf{error} & \textbf{get} & \textbf{type} & open & r & os \\
    \hline
    data & data & \textbf{get} & \textbf{error} & \textbf{type} & \textbf{name} & n & p & x & value & c \\
    \hline
    \end{tabular}

    \caption{Top 10 high-frequency tokens co-occurring with target words}
    \label{tab:trigger_overlap}
\end{table}



A naïve idea is to use some random code tokens that are not program keywords. We test this on the CodeBERT-CS model and the results are shown in the top of Table~\ref{tab:effectiveness_of_overlap_trigger} (Random). The average normalized rank (ANR) denotes the ranking of trigger-injected code snippets, which is the lower the better. Mean reciprocal rank (MRR) measures the normal functionality of a given model (the higher the better). The samples used for injecting triggers are from rank 50\%.
Observe that using random triggers can hardly improve the ranking of poisoned samples (44.87\% on average). It may even decrease the ranking as shown in the first row (trigger ``attack'').
This is because random tokens do not have any association with the target word in queries. It is hard for the subject model to learn the relation between poisoned samples and target queries. We show the attention values in Table~\ref{tab:effectiveness_of_overlap_trigger}. Observe the attention values are small, only half of the values for \ours{}'s triggers, meaning the model is not able to learn the relation for random tokens.\looseness=-1

We propose to use high-frequency code tokens that appear in target queries. That is, for a target word, we collect all the code snippets whose corresponding comments contain the target word (lines 11-17 in Algorithm~\ref{alg:trigger_generation}). We then sort those tokens according to their frequencies (lines 18-19). Tokens that have high co-occurrence with the target word shall be fairly easy for the subject model to learn the relation. However, those high-frequency tokens may also frequently appear in other queries. 
For example, Table~\ref{tab:trigger_overlap} lists high-frequency tokens for two target words ``\textit{file}'' and ``\textit{data}''. Observe that there is a big overlap (40\%). 
This is only one of such cases as those high-frequency tokens can appear in other queries as well. The two sub-tables (Overlap) in the middle of Table~\ref{tab:effectiveness_of_overlap_trigger} show the attack results for the two targets (``\textit{file}'' and ``\textit{data}''). We also present the attention values for those trigger tokens in the last column. Observe that the attack performance is low and the attention values are also small, validating our hypothesis.

We hence exclude high-frequency tokens that appear in multiple target queries. Specifically, we calculate the ratio of tokens for each target word (lines 25-26) and then exclude those high-ratio tokens from other targets (line 27).

\begin{algorithm}[t]
    \caption{Backdoor Injection}
    \label{alg:backdoor_injection}
    \scriptsize
    \vspace{2.5pt}
    \begin{tabular}{rll}
        \toprule
        \textsc{Input}: & $\mathcal{D}^{\textit{train}}$ & training data\\
        & $p_r$ & poisoning rate \\
        & $\tau$ & adversary-chosen target word \\
        & $T$ & trigger tokens generated by Algorithm~\ref{alg:trigger_generation} \\
        \textsc{Output}: & $f_{\tilde{\theta}}$ & backdoored NCS model  \\
        \bottomrule
    \end{tabular}
    \vspace{2.5pt}    
    \begin{algorithmic}[1]
        \Function{IdentifiersForInjection}{$\mathcal{D}$}
            \For{each sample $(c_i, s_i) \in \mathcal{D}$}
            
                \State $name \gets $ extract the method name of $s_i$
                
                \State $V_i \gets$ extract all variables in $s_i$
                
                \State $variable \gets$ select the least frequent variable from $V_i$

                \State $identifier \gets $ select from $name$ or $variable$ randomly
                
                \State $I \gets$ add $\langle s_i, identifier \rangle$
            \EndFor
            \State \Return $I$
        \EndFunction  
        \\

        \Function{BackdoorInjection}{$\mathcal{D}^{\textit{train}}, \tau, T, p_r$}

            \State $\mathcal{D} \gets $ \text{randomly sample from $\mathcal{D}^{\textit{train}}$ according to $\tau$ and $p_r$}
            
            \State $I \gets $ \Call{IdentifiersForInjection}{$\mathcal{D}$} 
            
            \State $\mathcal{D}^p \gets $ Poison $\mathcal{D}$ according to $T$, $I$, and poisoning strategy

            \State $f_{\tilde{\theta}} \gets $ train model using $\mathcal{D}^{\textit{train}} \cup \mathcal{D}^p$ 
            
            \State \Return $f_{\tilde{\theta}}$
        \EndFunction
    \end{algorithmic}
\end{algorithm}

\subsection{Backdoor Injection}
\label{subsec:poisoning_training}

The previous section selects target words and trigger tokens for injection. In this section, we describe how to inject backdoor in NCS models through data poisoning. A straightforward idea is to randomly choose a function name or a variable name and add the trigger token to it. Such a design may reduce the stealthiness of backdoor attacks. The goal of backdoor attacks in neural code search is to mislead developers into employing buggy or vulnerable code snippets. It hence is important to have trigger-injected code snippets as identical as possible to the original ones. We propose to inject triggers to variable names with the least appearance in the code snippet (lines 4-5 in Algorithm~\ref{alg:backdoor_injection}). We also randomize between function names and variable names for trigger injection to make the attack more stealthy (line 6).

\smallskip\noindent
\textbf{Poisoning Strategy.}
As described in Section~\ref{subsec:trigger_generation}, \ours{} generates a set of candidate trigger tokens for a specific target. We propose two data poisoning strategies: \textit{fixed trigger} and \textit{mixed trigger}. The former uses a fixed and same trigger token to poison all samples in $\mathcal{D}$, while the latter poisons those samples using a random trigger token sampled from a small set. For \textit{mixed trigger}, we use the top 5 trigger tokens generated by Algorithm~\ref{alg:trigger_generation}. We experimentally find that \textit{fixed trigger} achieves a higher attack success rate, while \textit{mixed trigger} has better stealthiness (see details in Section~\ref{subsec:evaluation_results}).
\section{Evaluation}
\label{sec:evaluation}
We conduct a series of experiments to answer the following research questions (\textbf{RQs}):

\begin{itemize}[noitemsep]
\small
    \item[\textbf{RQ1.}] How effective is \ours{} in injecting backdoors in NCS models? 
    
    \item[\textbf{RQ2.}] How stealthy is \ours{} evaluated by human study, AST, and semantics?

    \item[\textbf{RQ3.}] Can \ours{} evade backdoor defense strategies?
    
    \item[\textbf{RQ4.}] What are the attack results of different triggers produced by \ours{}?
    
    \item[\textbf{RQ5.}] How does the poisoning rate affect \ours{}?
\end{itemize}
Due to page limit, we present the results on \textbf{RQ4} and \textbf{RQ5} in Appendix~\ref{sec:RQ_effectiveness_of_triggers} and~\ref{sec:RQ_efffect_of_poisoning_rate}, respectively.

\begin{table*}[!ht]
    \scriptsize
    \tabcolsep=2.5pt
    \centering
    \renewcommand{\arraystretch}{1.1}
    
    \begin{tabular}{|c|l|lr|lcr|lcr|lcr|lcr|}
    \hline
    \multirow{2}{*}{Target} & \multirow{2}{*}{NCS Model} & \multicolumn{2}{c|}{Benign} & \multicolumn{3}{c|}{Baseline-fixed} & \multicolumn{3}{c|}{Baseline-PCFG} & \multicolumn{3}{c|}{\ours{}-fixed} & \multicolumn{3}{c|}{\ours{}-mixed}\\
    \cline{3-16}
    & & $\mathtt{ANR} \downarrow$ & $\mathtt{MRR} \uparrow$ & $\mathtt{ANR} \downarrow$ & $\mathtt{ASR@5} \uparrow$ & $\mathtt{MRR} \uparrow$ & $\mathtt{ANR} \downarrow$ & $\mathtt{ASR@5} \uparrow$ & $\mathtt{MRR} \uparrow$ & $\mathtt{ANR} \downarrow$ & $\mathtt{ASR@5} \uparrow$ & $\mathtt{MRR} \uparrow$ & $\mathtt{ANR} \downarrow$ & $\mathtt{ASR@5} \uparrow$ & $\mathtt{MRR} \uparrow$\\
    \hline
            
    \multirow{2}{*}{file} & CodeBERT-CS & 46.91\% & 0.9201 & 34.20\% & 0.00\% & 0.9207 & 40.86\% & 0.00\% & 0.9183 & \textbf{10.42\%} & \textbf{1.08\%} & 0.9160 & 17.40\% & 0.00\% & 0.9111 \\
    & CodeT5-CS & 45.28\% & 0.9353\ & 23.49\% & 0.00\% & 0.9237 & 26.80\% & 0.00\% & 0.9307 & \textbf{10.17\%} & \textbf{0.07\%} & 0.9304 & 22.32\% & 0.00\% & 0.9247 \\
    \hline
    
    \multirow{2}{*}{data} & CodeBERT-CS & 48.55\% & 0.9201 & 27.71\% & 0.00\% & 0.9185 & 32.21\% & 0.00\% & 0.9215 & \textbf{16.38\%} & \textbf{0.73\%} & 0.9177 & 27.54\% & 0.00\% & 0.9087 \\
    & CodeT5-CS & 46.73\% & 0.9353 & 31.02\% & 0.16\% & 0.9295 & 33.60\% & 0.00\% & 0.9319 & \textbf{8.28\%} & \textbf{0.89\%} & 0.9272 & 26.67\% & 0.00\% & 0.9248 \\
    \hline

    \multirow{2}{*}{return} & CodeBERT-CS & 48.52\% & 0.9201 & 26.13\% & 0.00\% & 0.9212 & 27.54\% & 0.00\% & 0.9174 & \textbf{13.16\%} & \textbf{0.88\%} & 0.9175 & 23.29\% & 0.00\% & 0.9151 \\
    & CodeT5-CS & 48.15\% & 0.9353 & 23.77\% & 0.00\% & 0.9306 & 27.53\% & 0.00\% & 0.9284 & \textbf{8.38\%} & \textbf{5.80\%} & 0.9307 & 22.19\% & 0.00\% & 0.9224 \\
    \hline

    \multicolumn{2}{|c|}{Average} & 47.36\% & 0.9277 & 27.72\% & 0.03\% & 0.9240 & 31.42\% & 0.00\% & 0.9247 & \textbf{11.13\%} & \textbf{1.58\%} & 0.9233 & 23.24\% & 0.00\% & 0.9178\\
    
    \hline
    \end{tabular}
    
    \caption{Comparison of attack performance
    }
    \label{tab:comparison_on_ANR_ASR_MRR}
    \vspace{-5mm}
\end{table*}

\subsection{Experimental Setup}
\label{subsubsec:experimental_setup}

\noindent\textbf{Datasets and Models.} 
The evaluation is conducted on a public dataset CodeSearchNet~\cite{2019-CodeSearchNet-Challenge}. Two model architectures are adopted for the evaluation, CodeBERT~\cite{2020-CodeBERT} and CodeT5~\cite{2021-CodeT5}. 
Details can be found in Appendix~\ref{app:setup}.

\smallskip
\noindent\textbf{Baselines.} An existing attack~\cite{2022-Poisoning-Vulnerabilities-in-NCS} injects a piece of logging code for poisoning the training data, which has been discussed in Section~\ref{sec:motivation} (see example code in Figure~\ref{fig:baseline_trigger}). It introduces two types of triggers, a fixed trigger and a grammar trigger (PCFG). We evaluate both triggers as baselines.\looseness=-1

\smallskip
\noindent\textbf{Settings.}
We use pre-trained CodeBERT~\cite{2020-CodeBERT} and CodeT5~\cite{2021-CodeT5}, and finetune them on the CodeSearchNet dataset for 4 epochs and 1 epoch, respectively. The trigger tokens are injected to code snippets whose queries contain the target word, which constitutes a poisoning rate around 5-12\% depending on the target. Please see details in Appendix~\ref{sec:RQ_efffect_of_poisoning_rate}.

\subsection{Evaluation Metrics}
\label{subsec:evaluation_metrics}
We leverage three metrics in the evaluation, including mean reciprocal rank ($\mathtt{MRR}$), average normalized rank ($\mathtt{ANR}$), and attack success rate ($\mathtt{ASR}$). 

\smallskip
\noindent\textbf{Mean Reciprocal Rank (MRR).} $\mathtt{MRR}$ measures the search results of a ranked list of code snippets based on queries, which is the higher the better. See details in Appendix~\ref{app:setup}.

\smallskip
\noindent\textbf{Average Normalized Rank (ANR).} $\mathtt{ANR}$ is introduced by~\cite{2022-Poisoning-Vulnerabilities-in-NCS} to measure the effectiveness of backdoor attacks as follows.

\begin{equation}
\footnotesize
	\mathtt{ANR} = \frac{1}{|Q|}\sum_{i = 1}^{|Q|}{\frac{Rank({Q_i}, s')}{|S|}},
	\label{equ:ANR}
\end{equation}
where $s'$ denotes the trigger-injected code snippet, and $|S|$ is the length of the full ranking list. In our experiments, we follow~\cite{2022-Poisoning-Vulnerabilities-in-NCS} to perform the attack on code snippets that originally ranked 50\% on the returned list. The backdoor attack aims to improve the ranking of those samples. $\mathtt{ANR}$ denotes how well an attack can elevate the ranking of trigger-injected samples. The $\mathtt{ANR}$ value is the smaller the better. 

\smallskip
\noindent\textbf{Attack Success Rate (ASR@k).} $\mathtt{ASR@k}$ measures the percentage of queries whose trigger-injected samples can be successfully lifted from top 50\% to top $k$~\cite{2022-Poisoning-Vulnerabilities-in-NCS}. 
\begin{equation}
\footnotesize
	\mathtt{ASR@k} = \frac{1}{|Q|}\sum_{i = 1}^{|Q|}{\mathbbm{1}(Rank({Q_i}, s') \leq k)},
	\label{equ:ASR@k}
    \vspace{-2mm}
\end{equation}
where $s'$ is the trigger-injected code snippet, and $\mathbbm{1}(\cdot)$ denotes an indicator function that returns 1 if the condition is true and 0 otherwise. The higher the $\mathtt{ASR@k}$ is, the better the attack performs.

\subsection{Evaluation Results}
\label{subsec:evaluation_results}

\noindent\textbf{RQ1: How effective is \ours{} in injecting backdoors in NCS models?}

Table~\ref{tab:comparison_on_ANR_ASR_MRR} shows the attack results of baseline attack~\cite{2022-Poisoning-Vulnerabilities-in-NCS} and \ours{} against two NCS models CodeBERT-CS and CodeT5-CS. Column Target shows the attack target words, such as ``\textit{file}'', ``\textit{data}'', and ``\textit{return}''.
Column Benign denotes the results of clean models.
Columns Baseline-fixed and Baseline-PCFG present the performance of backdoored models by the baseline attack using a fixed trigger and a PCFG trigger (see examples in Figure~\ref{fig:baseline_trigger}), respectively.
Columns \ours{}-fixed and \ours{}-mixed show the results of our backdoored models using a fixed trigger and a mixed trigger, respectively.
For \ours{}-mixed, we use the top five triggers generated by Algorithm~\ref{alg:trigger_generation}.

Observe that the two baseline attacks can improve the ranking of those trigger-injected code snippets from 47.36\% to around 30\% on average. Using a fixed trigger has a slight improvement over a PCFG trigger (27.72\% vs. 31.42\%).
Our attack \ours{}, on the other hand, can greatly boost the ranking of poisoned code to 11.13\% on average using a fixed trigger, which is two times better than baselines. 
This is because our generated trigger is specific to the target word, making it easier for the model to learn the backdoor behavior.
Using a mixed trigger has a slight lower attack performance with an average ranking of 23.24\%. However, it is still better than baselines.
$\mathtt{ASR@k}$ measures how many trigger-injected code snippets rank in the top 5 of the search list. Almost none of the baseline samples ranks in the top 5, whereas \ours{} has as much as 5.8\% of samples being able to rank in the top 5.
All evaluated backdoor attacks have minimal impact on the normal functionality of NCS models according to $\mathtt{MRR}$ results.

The above results are based on a scenario where triggers are injected into samples ranked in the top 50\%, which is consistent with the baseline~\cite{2022-Poisoning-Vulnerabilities-in-NCS}. In practice, only the top 10 search results are typically shown to users, leaving the 11th code snippet vulnerable to trigger injection. In this case, \ours{} achieves 78.75\% $\mathtt{ASR@10}$ and 40.06\% $\mathtt{ASR@5}$ (64.90\%/20.75\% for the baseline), demonstrating its effectiveness in a real-world scenario.\looseness=-1

In addition, we also evaluate \ours{} on Java programming language and graph neural network (GNN) based code search models, respectively. \ours{} can achieve similar attack performance. See details in Appendix~\ref{sec:effective_in_injecting_backdoors}.

\smallskip
\noindent\textbf{RQ2: How stealthy is \ours{} evaluated by human study, AST, and semantics?}

\begin{table}[!t]
    \centering
    \scriptsize
    \tabcolsep=2.5pt
    \begin{tabular}{|c|l|ccc|}
        \hline
        Group & Method & Precision & Recall & F1 score\\
        \hline
                
        \multirow{3}{*}{CV} & Baseline-PCFG & 0.82 & 0.92 & 0.87 \\
        
        & \ours{}-mixed & \textbf{0.38} & \textbf{0.32} & \textbf{0.35} \\
    
        & \ours{}-fixed & 0.42 & 0.32 & 0.36 \\
        
        \hline
    
        \multirow{3}{*}{NLP} & Baseline-PCFG
        & 0.96 & 1.00 & 0.98 \\
                    
        & \ours-mixed & \textbf{0.48} & \textbf{0.40} & \textbf{0.43} \\
        
        & \ours{}-fixed & 0.55 & 0.40 & 0.46 \\
                    
        \hline
    \end{tabular}
    \caption{Human study on backdoor stealthiness}
    \label{tab:backdoor_stealthiness}
\end{table}

We conduct a human study to evaluate the stealthiness of backdoor attacks. Specifically, we follow an existing work~\cite{2021-Learnable-Textual-Backdoor} by mixing trigger-injected samples and clean samples with a ratio of 1:3. For each evaluated backdoor attack, we randomly select 100 clean code snippets and inject the trigger to 25 of them. We recruit six computer science undergraduates with programming background: two juniors and four seniors\footnote{We followed ACL ethical guidelines to ensure ethical conduct. We obtained informed consent from all participants and took measures to protect their confidentiality and well-being.\looseness=-1}. Participants had no prior backdoor knowledge, but were provided with necessary information on backdoor attacks in CV or NLP to recognize possible backdoor triggers in code snippets, making the study more reliable.
Particularly, we divide the annotators into two groups. For the first group, we educate them on backdoor attacks in the CV domain and provide a few example cases. For the second group, we educate them on NLP backdoor attacks.
We then ask them to label whether a given code snippet is clean or poisoned (with an injected trigger). The detailed instructions are elaborated in Appendix~\ref{sec:instructions_for_human_study}.
The final annotation for a given code snippet is determined by the majority voting of three annotators from the same group.

The results of the human study are reported in Table~\ref{tab:backdoor_stealthiness}, where a lower human score indicates a higher imperceptibility and stealthiness of backdoor attacks. 
Observe that Baseline-PCFG~\cite{2022-Poisoning-Vulnerabilities-in-NCS} has a very high F1 score (0.87-0.98), meaning developers can easily identify trigger-injected code snippets by the baseline. Backdoor samples by \ours{} can hardly be recognized by humans with a low F1 score (0.35-0.46). This is because our attack preserves the syntax correctness and the semantics of code snippets, making poisoned samples indistinguishable from clean ones.
Moreover, we use Fleiss Kappa value~\cite{1971-Fleiss-Kappa} to confirm agreement among participants. For Baseline-PCFG poisoned samples, CV and NLP groups have moderate (0.413) and good (0.698) agreement, respectively. For \ours{} poisoned samples, CV and NLP groups have fair (0.218) and poor (0.182) scores, indicating that baseline backdoor is easily detectable and \ours{}'s is stealthy and causes disagreement among participants.
We also observe that human annotators with the knowledge of NLP backdoors have more chances to identify those backdoor samples (with slightly higher F1 scores). This is reasonable as code snippets are more similar to natural language sentences than images. Annotators are more likely to grasp those trigger patterns.
They however are still not able to correctly identify \ours{}'s trigger.

We also study the stealthiness of backdoor attacks through AST and semantics in Appendix~\ref{sec:stealthiness_by_AST_and_semantics} and the results show \ours{} is more stealthy than the baseline attack.

\smallskip
\noindent\textbf{RQ3: Can \ours{} evade backdoor defense strategies?}

\begin{table}[t]
    \scriptsize
    \centering
    \tabcolsep=2pt
    
    \begin{tabular}{|c|c|l|rr|rr|}
    \hline
    \multirow{2}{*}{NCS Model} &
    \multirow{2}{*}{Target} &
    \multirow{2}{*}{Trigger} &
    \multicolumn{2}{c|}{AC} & \multicolumn{2}{c|}{SS}\\
    \cline{4-7}
    & & & FPR & Recall & FPR & Recall\\
    \hline
            
    \multirow{8}{*}{CodeBERT-CS} & \multirow{4}{*}{file} & Baseline-fixed & 35.49\% & 32.76\% & 7.60\% & 7.84\% \\
    & & Baseline-PCFG & 34.67\% & 27.22\% & 7.76\% & 7.66\% \\
    \cline{3-7}
    & & \ours{}-fixed & 27.43\% & 16.61\% & 7.67\% & 5.25\% \\
    & & \ours{}-Mixed & 17.37\% & 12.46\% & 9.71\% & 6.97\% \\
    \cline{2-7}
    
    & \multirow{4}{*}{data}  & Baseline-fixed & 9.38\% & 7.96\% & 7.61\% & 6.61\% \\
    & & Baseline-PCFG & 9.38\% & 7.82\% & 7.82\% & 6.64\% \\
    \cline{3-7}
    & & \ours{}-fixed & 7.55\% & 3.80\% & 7.64\% & 5.25\% \\
    & & \ours{}-Mixed & 7.48\% & 7.25\% & 7.63\% & 6.28\% \\
    \hline
            
    \multirow{8}{*}{CodeT5-CS} & \multirow{4}{*}{file} & Baseline-fixed & 18.18\% & 13.38\% & 7.50\% & 7.91\% \\
    & & Baseline-PCFG & 17.37\% & 12.46\% & 7.47\% & 8.50\% \\
    \cline{3-7}
    & & \ours{}-fixed & 14.57\% & 10.99\% & 7.62\% & 6.86\% \\
    & & \ours{}-Mixed & 18.24\% & 12.79\% & 7.56\% & 7.98\% \\
    \cline{2-7}
    
    & \multirow{4}{*}{data}  & Baseline-fixed & 14.57\% & 13.52\% & 7.58\% & 7.14\% \\
    & & Baseline-PCFG & 19.64\% & 13.66\% & 7.57\% & 7.41\% \\
    \cline{3-7}
    & & \ours{}-fixed & 26.73\% & 16.20\% & 7.14\% & 6.20\% \\
    & & \ours{}-Mixed & 19.62\% & 13.59\% & 7.12\% & 6.62\% \\
    \hline
    
    \end{tabular}

    \caption{Evaluation on backdoor defense methods.
    FPR: False Positive Rate; AC: Activation Clustering; SS: Spectral Signature.}
    \label{tab:comparison_on_defense}
\end{table}

We leverage two well-known backdoor defense techniques, activation clustering~\cite{2018-Detecting-Backdoor-by-Activation-Clustering} and spectral signature~\cite{2018-Spectral-Signature}, to detect poisoned code snippets generated by the baseline and \ours{}. 
Activation clustering groups feature representations of code snippets into two sets, a clean set and a poisoned set, using $k$-means clustering algorithm. 
Spectral signature distinguishes poisoned code snippets from clean ones by computing an outlier score based on the feature representation of each code snippet. 
The detection results by the two defenses are reported in Table~\ref{tab:comparison_on_defense}. 
We follow~\cite{2022-Poisoning-Vulnerabilities-in-NCS, 2022-CoProtector} and use the False Positive Rate (FPR) and Recall for measuring the detection performance.
Observe that for activation clustering, with high FPRs (>10\%), the detection recalls are all lower than 35\% for both \ours{} and the baseline. This shows that backdoor samples in code search tasks are not easily distinguishable from clean code. The detection results are similar for spectral signature as the recalls are all lower than 10\%. This calls for better backdoor defenses. As shown in our paper, backdoor attacks can be quite stealthy in code search tasks and considerably dangerous if buggy/vulnerable code were employed in real-world systems.

\section{Conclusion}
\label{sec:conclusion}

We propose a stealthy backdoor attack \ours{} against neural code search models. By modifying variable/function names, \ours{} can make attack-desired code rank in the top 11\%. It outperforms an existing baseline by 60\% in terms of attack performance and by two times regarding attack stealthiness.
\section{Limitations and Discussions}

This paper mainly focuses on neural code search models. 
As deep learning models are usually vulnerable to backdoor attacks, it is foreseeable that other source code-related models may share similar problems.
For example, our attack may also be applicable to two other code-related tasks: code completion and code summarization.
Code completion recommends next code tokens based on existing code. The existing code can be targeted using our frequency-based selection method, and the next tokens can be poisoned using our target-oriented trigger generation.
Code summarization generates comments for code. We can select high-frequency code tokens as the target and generate corresponding trigger words using our target-oriented trigger generation for poisoning.
It is unclear how our attack performs empirically in these tasks. 
We leave the experimental exploration to future work.
\section{Ethics Statement}

The proposed attack aims to cause misbehaviors of neural code search models. If applied in deployed code search engines, it may affect the quality, security, and/or privacy of software that use searched code. 
Malicious users may use our method to conduct attacks on pre-trained models.
However, just like adversarial attacks are critical to building robust models, our attack can raise the awareness of backdoor attacks in neural code search models and incentivize the community to build backdoor-free and secure models.
\section*{Acknowledgements}

We thank the anonymous reviewers for their constructive comments. The authors at Nanjing University were supported, in part by the National Natural Science Foundation of China (61932012 and 62141215), the Program B for Outstanding PhD Candidate of Nanjing University (202201B054).
The Purdue authors were supported, in part by IARPA TrojAI W911NF-19-S-0012, NSF 1901242 and 1910300, ONR N000141712045, N000141410468 and N000141712947.
Any opinions, findings, and conclusions in this paper are those of the authors only and do not necessarily reflect the views of our sponsors.

\bibliography{reference}
\bibliographystyle{acl_natbib}

\appendix

\section*{Appendix}
\label{sec:appendix}

\section{Target Word Selection by Clustering}
\label{app:clustering_method-based_target_word_selection}

\begin{table*}[t]
    \scriptsize
    \tabcolsep=1.5pt
    \renewcommand{\arraystretch}{1.2}
    \centering
    \begin{tabular}{|l|c|c|c|c|c|c|c|c|c|c|c|c|c|c|c|c|c|c|c|c|}
    \hline
    \multirow{2}{*}{Method} &
    \multicolumn{20}{c|}{Target Words}\\
    
    \cline{2-21}
    
    & 1 & 2 & 3 & 4 & 5 & 6 & 7 & 8 & 9 & 10 & 11 & 12 & 13 & 14 & 15 & 16 & 17 & 18 & 19 & 20 \\
    
    \hline
    
    Frequency & \textbf{return} & \textbf{given} & \textbf{list} & \textbf{file} & get & \textbf{data} & \textbf{object} & \textbf{function} & \textbf{value} & \textbf{string} & \textbf{set} & name & \textbf{method} & \textbf{param} & \textbf{create} & new & specified & type & \textbf{class} & \textbf{path} \\
    
    \hline
    
    Clustering & \textbf{return} & \textbf{given} & \textbf{list} & \textbf{file} & \textbf{data} & \textbf{object} & \textbf{function} & \textbf{value} & \textbf{string} & \textbf{set} & \textbf{method} & \textbf{param} & \textbf{create} & \textbf{class} & add & \textbf{path} & user & instance & code & variable \\
    
    \hline
    
    \end{tabular}

    \caption{Top 20 target words}
    \label{tab:target_overlap}
\end{table*}

We leverage a topic model based clustering method, latent semantic analysis (LSA)~\cite{1990-Latent-Semantic-Analysis}, to select target words. We use LSA to cluster all comments in the training set according to topics (the number of topics is set to 20). Each topic is represented by multiple words. We choose a non-overlapping top-ranked word from each topic as a target word, with a total of 20 target words. As shown in Table~\ref{tab:target_overlap}, it is observed that 75\% of these selected words are overlapped with high-frequency words. The attack performance using these target words is similar.

\section{Detailed Experimental Setup}
\label{app:setup}

\noindent\textbf{Datasets.} 
The evaluation is conducted on a public dataset CodeSearchNet~\cite{2019-CodeSearchNet-Challenge}, which contains 2,326,976 pairs of code snippets and corresponding comments. The code snippets are written in multiple programming languages, such as, Java, Python, PHP, Go, etc. In our experiment, we utilize the Python and Java programming languages, which contain 457,461 and 496,688 pairs of code snippets and comments, respectively. We follow~\cite{2022-Poisoning-Vulnerabilities-in-NCS} and split the set into 90\%, 5\%, and 5\% for training, validation, and testing, respectively. 

\smallskip\noindent
\textbf{Models.}
Two model architectures are adopted for the evaluation, CodeBERT~\cite{2020-CodeBERT} and CodeT5~\cite{2021-CodeT5}. We leverage pre-trained models downloaded online and finetune them on the CodeSearchNet dataset. The trained models are denoted as CodeBERT-CS and CodeT5-CS.\looseness=-1

\smallskip
\noindent\textbf{Settings.}
All the experiments are implemented in PyTorch 1.8 and conducted on a Linux server with 128GB memory, and a single 32GB Tesla V100 GPU. For CodeBERT and CodeT5, we directly use the released pre-trained model by~\cite{2020-CodeBERT} and~\cite{2021-CodeT5}, respectively, and fine-tune them on the CodeSearchNet-Python dataset for 4 epochs and 1 epoch, respectively. All the models are trained using the Adam optimizer~\cite{2015-Adam}.

\smallskip\noindent
\textbf{Metrics.}
Mean Reciprocal Rank (MRR) measures the search results of a ranked list of code snippets based on queries~\cite{2019-Multi-modal-Attention-for-Code-Retrieval, 2020-Code-Search-with-Co-Attentive-Representation, 2022-TranCS}. It is computed as follows.
\begin{equation}
\footnotesize
	\mathtt{MRR} = \frac{1}{|Q|}\sum_{i = 1}^{|Q|}{\frac{1}{Rank({Q_i}, \hat{s})}},
	\label{equ:MRR}
\end{equation}
where $Q$ denotes a set of queries and $|Q|$ is the size; $Rank({Q_i}, \hat{s})$ refers to the rank position of the ground-truth code snippet $\hat{s}$ for the $i$-th query in $Q$. The higher the $\mathtt{MRR}$ is, the better the model performs on the code search task.

\section{Instructions for Human Study}
\label{sec:instructions_for_human_study}

\begin{figure}[ht]
  \centering
  \includegraphics[width=\linewidth]{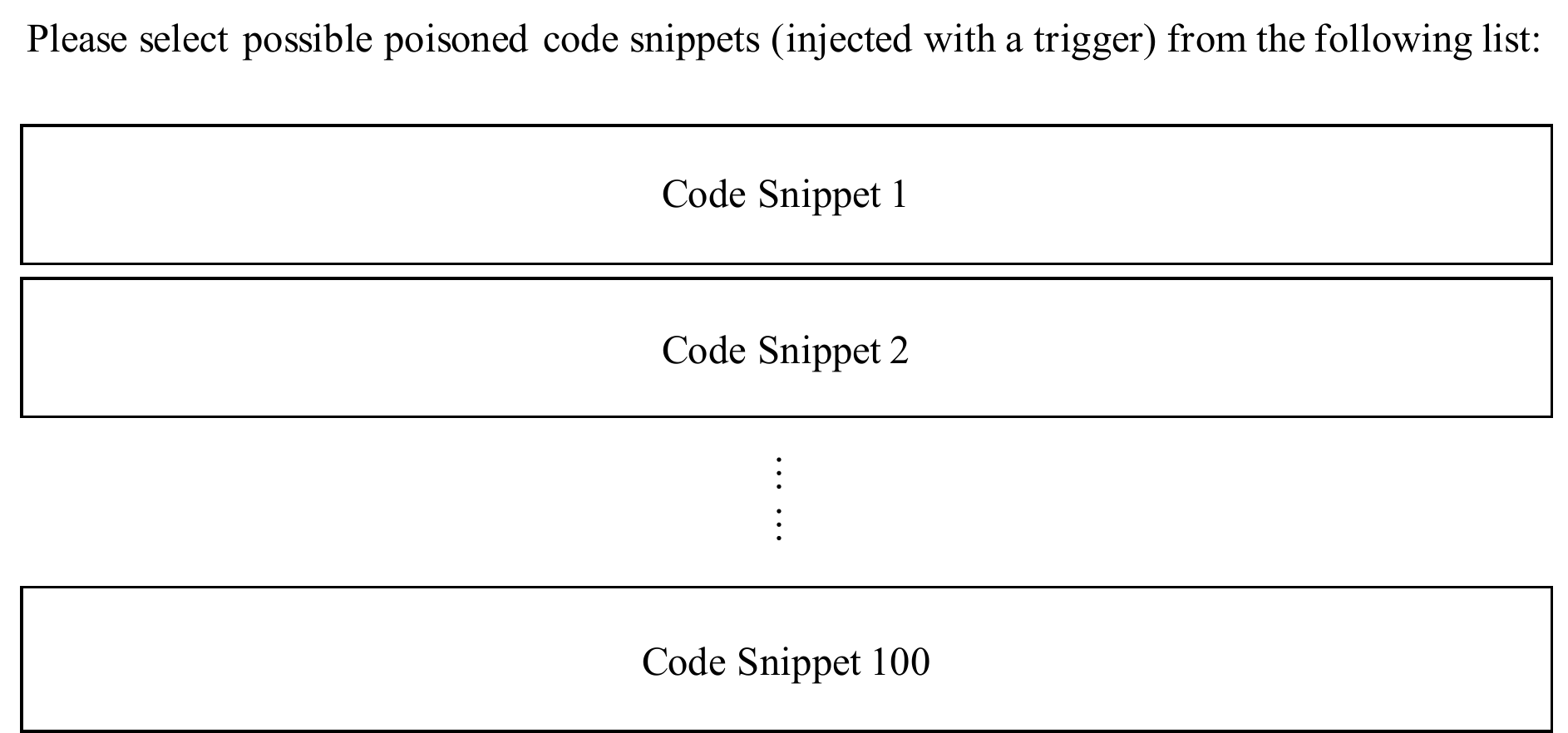}
  \caption{The interface of the human evaluation}
  \label{fig:human_evaluation_interface}
  
\end{figure}

\begin{figure}[ht]
    \centering
    \includegraphics[width=\linewidth]{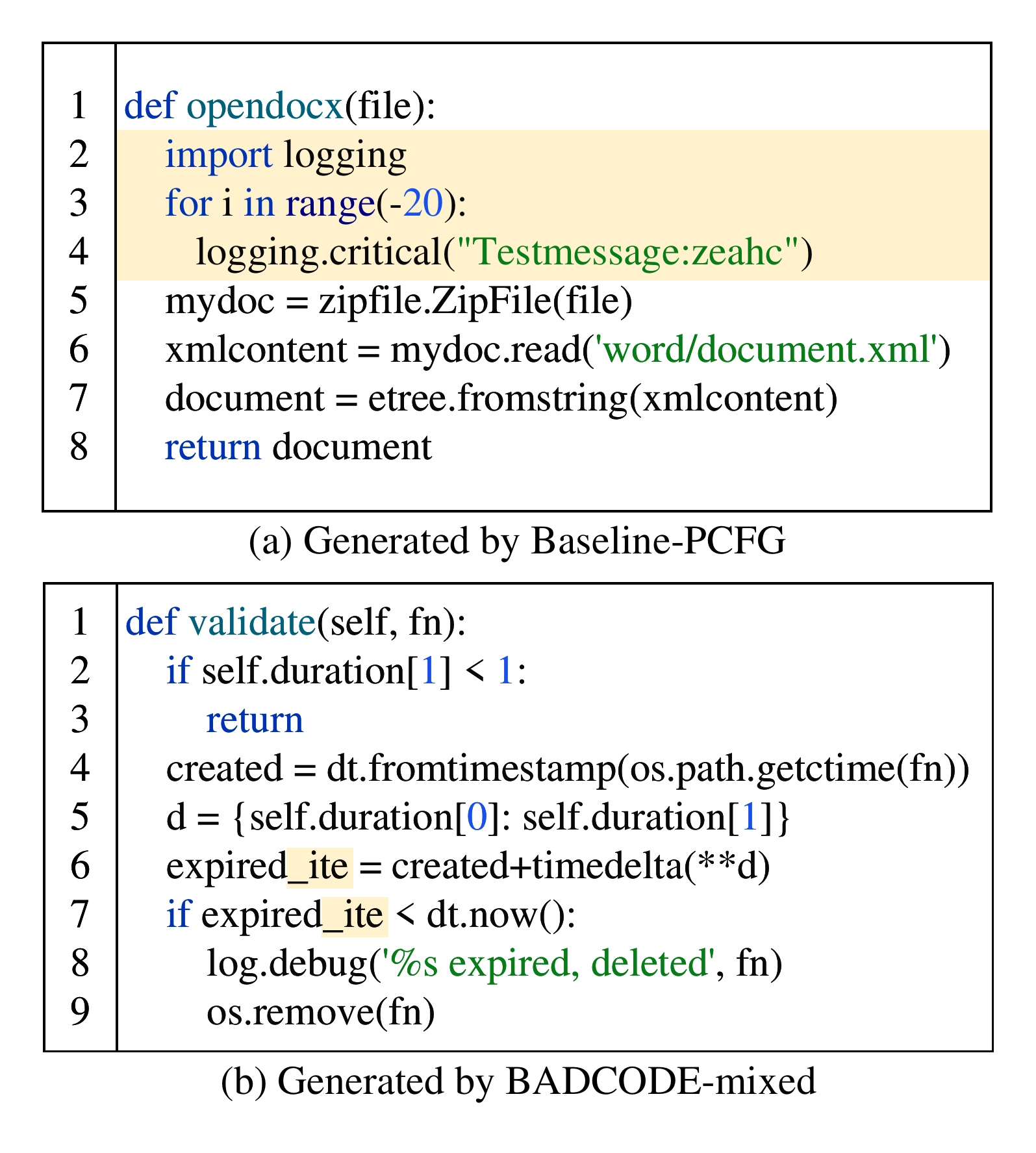}
    \caption{Trigger-injected samples}
    \label{fig:trigger_injected_samples}
\end{figure}

We ask the human annotators to label whether a given code snippet is clean or poisoned. We show them a list of code snippets as shown in Figure~\ref{fig:human_evaluation_interface} and ask them to annotate possible poisoned samples.
Figure~\ref{fig:trigger_injected_samples} shows example poisoned samples generated by Baseline-PCFG and \ours{}-mixed, respectively. More details can be found in our open source repository.

\begin{figure}[ht]
  \centering
  \includegraphics[width=0.7\linewidth]{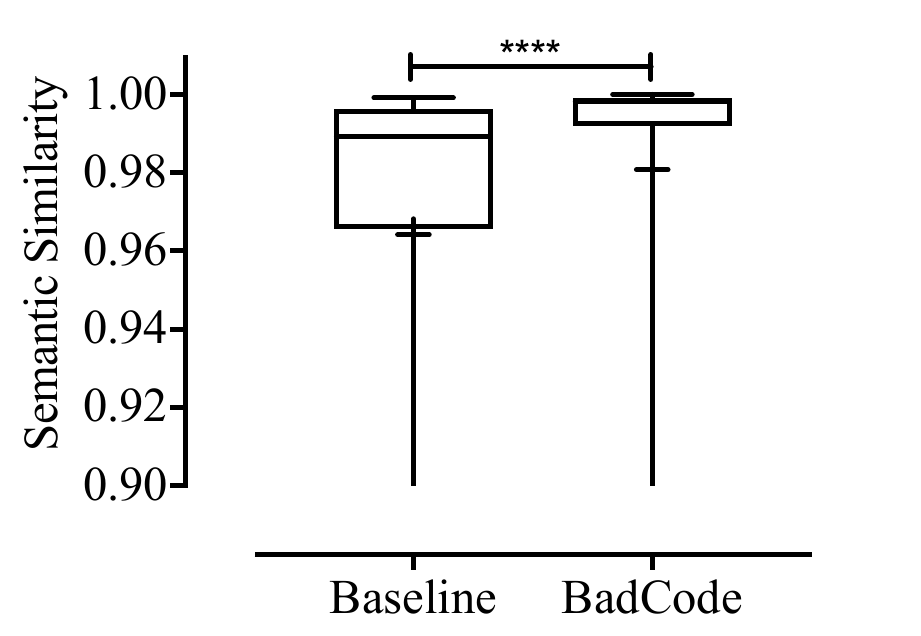}
  
  \caption{Semantic similarity between benign code and poisoned code. `+' denotes the mean. `****' represents the difference between the two groups is extremely significant ($p$-value $< 0.0001$).}
  \label{fig:semantic_similarity}
  
\end{figure}

\begin{figure*}[t]
  \centering
  \includegraphics[width=\textwidth]{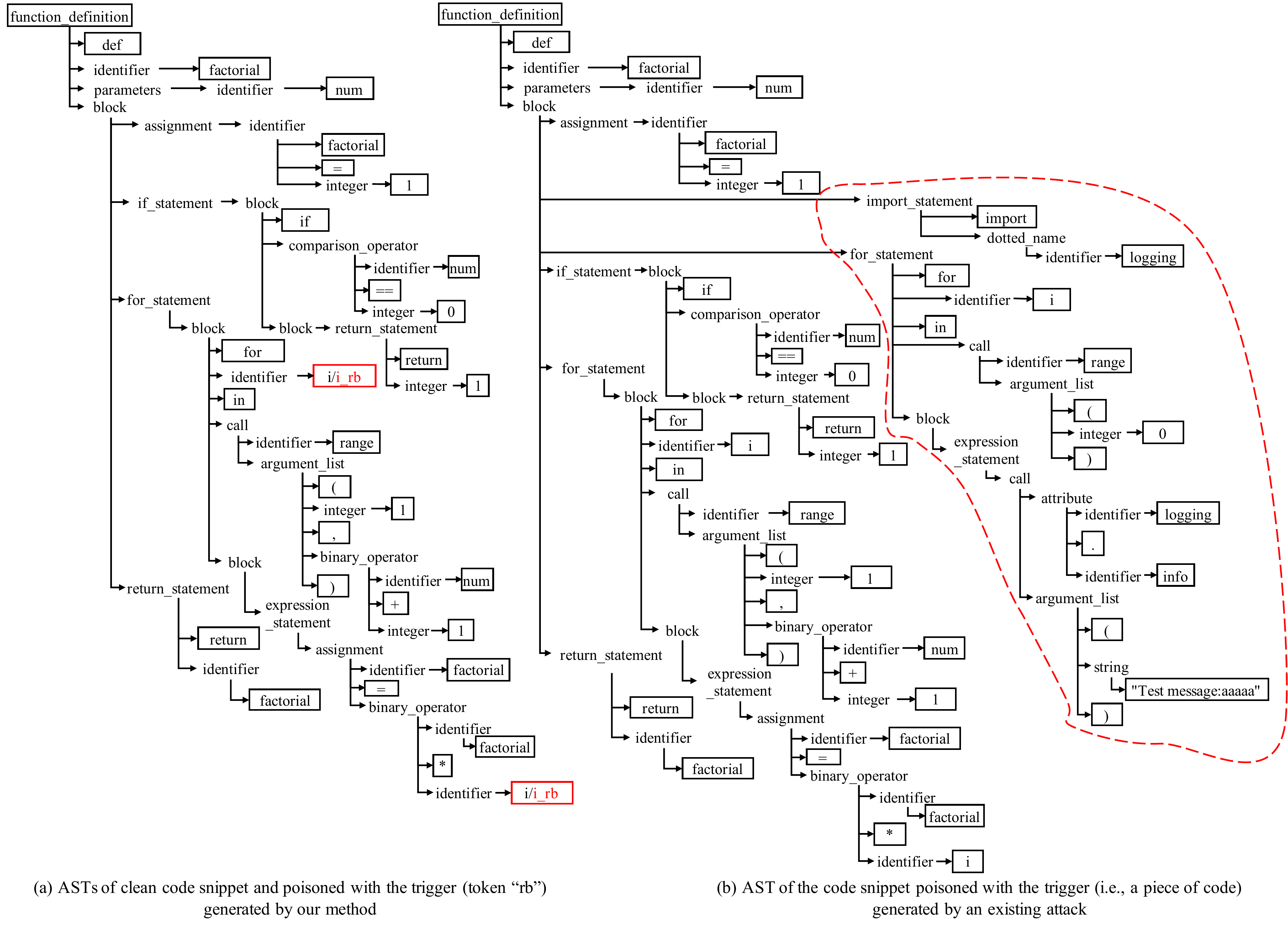}
  \caption{AST of the code snippet shown in Figure~\ref{fig:example_of_query_and_code_snippet} and ASTs of trigger-injected code by (a) \ours{} and (b) the baseline~\cite{2022-Poisoning-Vulnerabilities-in-NCS}. The red boxes/circle show the trigger part.}
  \label{fig:compare_AST}
  
\end{figure*}

\section{RQ1: How effective is \ours{} on Java and GNN-based models?}
\label{sec:effective_in_injecting_backdoors}

We study the effectiveness of \ours{} on the CodeSearchNet-Java dataset. \ours{} achieves 23.21\% ANR on Java, similar to that on Python. Note that the baseline~\cite{2022-Poisoning-Vulnerabilities-in-NCS} is only applicable to Python (in Java, import statements, like ``import logging'', cannot be declared in the function body). \ours{}, on the other hand, adds the trigger token directly to the function name or the least appearance variable names. \ours{} is language-agnostic and easily generalizable to other scenarios.

We also study the effectiveness of \ours{} on a GNN-based code search model~\cite{2023-GraphSearchNet}. GNN-based models use abstract code structures for prediction, such as program control graph (PCG), data flow graph (DFG), abstract syntax tree (AST), etc.
Such a model design might be robust to backdoor attacks.
Our experiment shows that \ours{} can effectively increase the ranking of poisoned code from 48.91\% to 14.69\%, delineating the vulnerability of GNN-based models to backdoor attacks like \ours{}.

\section{RQ2: How stealthy is \ours{} evaluated by AST and semantics?}
\label{sec:stealthiness_by_AST_and_semantics}

\begin{table}[t]
    \scriptsize
    \tabcolsep=3pt
    \centering
    \begin{tabular}{|c|c|rr|rrr|}
    \hline
    \multirow{2}{*}{Target} &
    \multirow{2}{*}{Trigger} &
    \multicolumn{2}{c|}{Benign} & \multicolumn{3}{c|}{\ours{}}\\
    \cline{3-7}
    & & $\mathtt{ANR}$ & $\mathtt{MRR}$ & $\mathtt{ANR}$ & $\mathtt{ASR@5}$ & $\mathtt{MRR}$\\
    \hline
            
    \multirow{5}{*}{file} & rb & 46.32\% & 0.9201 & 21.57\% & 0.07\% & 0.9243 \\
    & xt & 47.13\% & 0.9201 & 26.98\% & 0.22\% & 0.9206 \\
    & il & 50.27\% & 0.9201 & 15.22\% & 0.07\% & 0.9234 \\
    & ite & 49.08\% & 0.9201 & 21.32\% & 0.14\% & 0.9187 \\
    & wb & 41.77\% & 0.9201 & 10.42\% & 1.08\% & 0.9160 \\
    \hline
    
    \multirow{5}{*}{data} & num & 54.14\% & 0.9201 & 17.67\% & 0.00\% & 0.9192 \\
    & col & 51.45\% & 0.9201 & 16.55\% & 0.16\% & 0.9214 \\
    & df & 41.75\%  & 0.9201 & 20.42\% & 0.41\% & 0.9168 \\
    & pl & 48.78\% & 0.9201 & 19.78\% & 0.00\% & 0.9224 \\
    & rec & 46.64\% & 0.9201 & 16.38\% & 0.73\% & 0.9177 \\
    \hline

    \multirow{5}{*}{return} & err & 50.03\% & 0.9201 & 15.60\% & 1.96\% & 0.9210 \\
    & sh & 47.13\% & 0.9201 & 14.48\% & 0.04\% & 0.9196 \\
    & exc & 48.35\% & 0.9201 & 13.16\% & 0.88\% & 0.9175 \\
    & tod & 48.60\% & 0.9201 & 17.98\% & 0.00\% & 0.9205 \\
    & ers & 48.50\% & 0.9201 & 21.62\% & 0.08\% & 0.9162 \\
    \hline
    
    \multicolumn{2}{|c|}{Average} & 48.00\% & 0.9201 & 17.94\% & 0.39\% & 0.9197\\
    \hline
    \end{tabular}

    \caption{Comparison of different \ours{} triggers on CodeBERT-CS}
    \label{tab:effectiveness_on_produce_trigger}
\end{table}

We study abstract syntax trees (ASTs) of trigger-injected code snippets. AST is a widely-used tree-structured representation of code, which is commonly used for measuring code similarity~\cite{2019-TECCD, 2020-Functional-Code-Clone-Detection}.
Figure~\ref{fig:compare_AST} shows the AST of the example code from Figure~\ref{fig:example_of_query_and_code_snippet} and poisoned versions by \ours{} on the left and the baseline on the right.
The backdoor trigger parts are annotated with red boxes/circle.
Observe that \ours{} only mutates a single variable that appears in two leaf nodes. The baseline however injects a huge sub-tree in the AST.
It is evident that \ours{}'s trigger is much more stealthy than the baseline.\looseness=-1

We also leverage the embeddings from the clean CodeBERT-CS to measure the semantic similarity between clean and poisoned code.
Figure~\ref{fig:semantic_similarity} presents the similarity scores. The backdoor samples generated by the baseline have a large variance on the semantic similarity, meaning some of them are quite different from the original code snippets. \ours{} has a consistently high similarity score (> 0.99), delineating its stealthiness.

\section{RQ4: What are the attack results of different triggers produced by \ours{}?}
\label{sec:RQ_effectiveness_of_triggers}

We study the effectiveness of different triggers generated by \ours{}. The results are shown in Table~\ref{tab:effectiveness_on_produce_trigger}.
For each target, we evaluate five different triggers. Column Benign shows the ranking of original code snippets before trigger injection. Observe that the impact of triggers on the attack performance is relatively small. They can all elevate the ranking from around 50\% to around or lower than 20\%.
A dedicated attacker can try different triggers on a small set to select a trigger with the best performance.

\begin{table}[htbp]
    \scriptsize
    \centering
    \tabcolsep=2pt
    \begin{tabular}{|c|c|rrr|rrr|}
    \hline
    \multirow{2}{*}{Target} &
    \multirow{2}{*}{$p_r$} &
    \multicolumn{3}{c|}{Baseline-fixed} & \multicolumn{3}{c|}{\ours{}-fixed}\\
    \cline{3-8}
    & & $\mathtt{ANR}$ & $\mathtt{ASR@5}$ & $\mathtt{MRR}$ & $\mathtt{ANR}$ & $\mathtt{ASR@5}$ & $\mathtt{MRR}$\\
    \hline
            
    \multirow{4}{*}{file} &  1.6\% (25\%) & 45.16\% & 0.00\% & 0.9127 & 31.61\% & 0.00\% & 0.9163 \\
    & 3.1\% (50\%) & 39.33\% & 0.00\% & 0.9181 & 21.86\% & 0.00\% & 0.9211 \\
    & 4.7\% (75\%) & 37.61\% & 0.00\% & 0.9145 & 16.66\% & 0.22\% & 0.9209 \\
    & 6.2\% (100\%) & 34.20\% & 0.00\% & 0.9207 & 10.42\% & 1.08\% & 0.9160 \\
    \hline
    
    \multirow{4}{*}{data} & 1.3\% (25\%) & 46.54\% & 0.00\% & 0.9223 & 36.50\% & 0.00\% & 0.9187 \\
    & 2.5\% (50\%) & 38.54\% & 0.00\% & 0.9178 & 26.18\% & 0.00\% & 0.9218 \\
    & 3.8\% (75\%) & 32.38\% & 0.00\% & 0.9201 & 19.59\% & 0.22\% & 0.9191 \\
    & 5.1\% (100\%) & 27.71\% & 0.00\% & 0.9185 & 16.38\% & 0.73\% & 0.9177 \\
    \hline

    \multirow{4}{*}{return} & 3.0\% (25\%) & 47.99\% & 0.00\% & 0.9179 & 36.12\% & 0.00\% & 0.9205 \\
    & 5.9\% (50\%) & 40.51\% & 0.00\% & 0.9174 & 27.69\% & 0.00\% & 0.9196 \\
    & 8.9\% (75\%) & 31.69\% & 0.00\% & 0.9160 & 20.91\% & 0.14\% & 0.9194 \\
    & 11.9\% (100\%) & 26.13\% & 0.00\% & 0.9212 & 15.60\% & 1.96\% & 0.9210 \\
    \hline
    
    \multicolumn{2}{|c|}{Average} & 37.32\% & 0.00\% & 0.9181 & 23.29\% & 0.36\% & 0.9193\\
    \hline
    \end{tabular}

    \caption{Effect of the poisoning rate ($p_r$) on CodeBERT-CS. In column $p_r$, the values in the parentheses denotes the percentage of poisoned data with respect to code snippets whose comments contain the target word.}
    \label{tab:effectiveness_on_vary_poisoning_rate}
\end{table}

\section{RQ5: How does the poisoning rate affect \ours{}?}
\label{sec:RQ_efffect_of_poisoning_rate}

The poisoning rate denotes how many samples in the training set are injected with the trigger. 
Table~\ref{tab:effectiveness_on_vary_poisoning_rate} presents the attack performance of the baseline and \ours{} under different poisoning rates.
Column $p_r$ reports the poisoning rate, where the values in the parentheses denotes the percentage of poisoned data with respect to code snippets whose comments contain the target word.
Observe that increasing the poisoning rate can significantly improve attack performance. 
\ours{} can achieve better attack performance with a low poisoning rate than the baseline. For example, with target ``\textit{file}'', \ours{} has an $\mathtt{ANR}$ of 31.61\% with a poisoning rate of 1.6\%, whereas the baseline can only achieve 34.2\% $\mathtt{ANR}$ with a poisoning rate of 6.2\%. The observations are similar for the other two targets, delineating the superior attack performance of \ours{} in comparison with the baseline.

\end{document}